\documentclass[aps,nofootinbib,showpacs,preprintnumbers,amsmath,amssymb]{revtex4}

\def\be{\begin{equation}}
\def\ee{\end{equation}}
\def\ba{\begin{eqnarray}}
\def\ea{\end{eqnarray}}
\def \bea{\begin{eqnarray}}
\def \eea{\end{eqnarray}}

\usepackage{graphics}
\usepackage{graphicx}
\usepackage{dcolumn}
\usepackage{bm}
\usepackage{epsfig}
\usepackage{graphicx}
\usepackage{multirow}
\usepackage{color}
\usepackage{dcolumn}
\usepackage{graphicx,epsfig}%
\usepackage{subfig}
\def \ee{\end{equation}}
\def \be{\begin{equation}}
\def \bea{\begin{eqnarray}}
\def \eea{\end{eqnarray}}

\newcommand{\m}{\mu}
\newcommand{\n}{\nu}

\preprint{}
\begin{document}

\title{Black Hole Solutions for Scale Dependent Couplings:\\
The de Sitter and the Reissner-Nordstr\"om Case \\
}

\author{Benjamin Koch and Paola Rioseco}
 \affiliation{
Instituto de F\'{i}sica, \\
Pontificia Universidad Cat\'{o}lica de Chile, \\
Av. Vicu\~{n}a Mackenna 4860, \\
Santiago, Chile \\}
\date{\today}

\begin{abstract}
Allowing for scale dependence of the gravitational couplings
leads to a generalization of the corresponding field equations.
In this work, those equations are solved for the Einstein-Hilbert and the Einstein-Maxwell case,
leading to generalizations of the (Anti)-de Sitter and the Reissner-Nordstr\"om black holes.
Those solutions are discussed and compared to their classical counterparts.
\end{abstract}

\maketitle

%
\section{Introduction}

Black holes are very fascinating objects.
Apart from the fact that they are classical solutions
of the equations of general relativity, they also have additional ``features'', such as the existence of
an event horizon and an essential singularity at the origin ($r=0$, typically behind the event horizon).
The existence of this singularity can be interpreted as the break down of the 
validity of the classical theory that predicts such solutions.
In this sense the study of black holes gives an interesting opportunity
of exploring general relativity in the transition between a regime where
the classical solution is known to be valid to high precision, 
and a regime where corrections to the classical prediction are to be expected.
Those expected corrections to the classical regime
are broadly assumed to be of quantum nature.
A famous example for such quantum corrections is the predicted existence
of thermal radiation, emitted by massive black holes, and induced by
quantum fluctuations \cite{Hawking:1974rv}.

When one takes general relativity as classical limit of a quantum theory seriously 
one has to deal with a lot of conceptual and technical difficulties.
Although, those difficulties are far from settled, there exist various promising approaches
aiming for a full (or partial) quantum formulation of general relativity \cite{Wheeler:1957mu,Deser:1976eh,Rovelli:1997yv,Bombelli:1987aa,Ashtekar:2004vs,Sakharov:1967pk,Jacobson:1995ab,Verlinde:2010hp,Reuter:1996cp,Litim:2003vp,Horava:2009uw,Charmousis:2009tc,Ashtekar:1981sf,Penrose:1986ca,Connes:1996gi,Nicolini:2008aj,Gambini:2004vz}
(for a review see \cite{Kiefer:2005uk}), even
without the necessity of going to a very different theory in higher dimensions.
Although, the different approaches are quite diverse, most of them still have a common feature:
Like in most other quantum field theories, they show a non-trivial scale dependence.
This means that the effective couplings of the theories become actually scale dependent quantities.
For example, Newtons constant $G_0$ acquires a scale dependence $G_0\rightarrow G(k)$.
This scaling behavior is especially nice to se in the realizations of Weinbergs Asymptotic 
Safety program \cite{Weinberg:1979,Wetterich:1992yh,Dou:1997fg,Souma:1999at,Reuter:2001ag,Fischer:2006fz,Percacci:2007sz,Litim:2008tt}. Such a scaling is
expected to modify classical observables, such as the black hole background \cite{Bonanno:1998ye,Bonanno:2000ep,Bonanno:2006eu,Reuter:2006rg,Falls:2012nd,Falls:2010he,Cai:2010zh,Becker:2012js,Becker:2012jx,Koch:2013owa,Koch:2014cqa,Nojiri:1996vs}.
Even though, the way how this scale dependence is calculated and the functional form
of the scale dependence itself can vary among the different approaches, 
the pure existence of such a scale dependence seems to be a solid statement.

In this paper the idea of scale dependent couplings will be implemented
at the level of effective action and the corresponding improved equations of motion.
In order obtain more generic results, no particular functional form will be assumed
for the scale dependent couplings $(G(k), \dots)$. 
Instead, a symmetry criterion will be imposed on the metric ansatz, which actually allows
to solve the improved equations of motion without knowledge of the functional form
of the coupling constants $(G(k), \dots)$.
This technique will be used in order to obtain two new black hole solutions that are on the
one hand generalizations of the corresponding classical solution and on the other hand
solutions of the self-consistent field equations in the context of scale dependent couplings.
The corrections with respect to the classical solution 
will be parametrized in terms of an dimensionful integration constant $\epsilon$,
which for the case of quantum-gravity-induced scale dependence can be expected to be of the order
of the Planck mass $M_{Pl}$.

The paper is organized as follows.
In the subsection \ref{sec_RNclas} properties of the classical (Anti)-de Sitter ((A)dS) solution
and the classical Reissner-Nordstr\"om (RN) will be shortly summarized.
In the following subsection \ref{sec_scaleset} it will be reviewed how a scale dependence of the gravitational
couplings ($G(k), \dots$) can be implemented self-consistently at the level of 
generalized gravitational field equations.
In section \ref{sec_EHsol} the generalized (A)dS black hole solution for
the generalized Einstein-Hilbert equations will be discussed.
This solution is obtained and parametrized in subsection \ref{sec_findAdS},
then in subsection \ref{sec_AdSasympt} the asymptotic behavior of this solution is calculated.
A perturbative analysis of the thermodynamic properties of this solution is presented in
subsection \ref{sec_AdSthermo}.
In section \ref{sec_RNsol} the generalized RN black hole solution for the generalized Einstein-Maxwell
equations will be discussed.
The generalized RN solution is presented and parametrized in subsection \ref{sec_RNfind},
then in subsection \ref{sec_RNasym} the asymptotic behavior of this solution is calculated and
the conserved quantity corresponding to the electrical charge is calculated in subsection
\ref{sec_RNcharge}.
The horizon structure, the thermodynamic corrections, and
the cosmic censorship  of the solution are discussed in subsection
\ref{sec_RNtemp}.
The results of this work summarized in section \ref{sec_sum} and important features are highlighted.

\subsection{The classical (A)dS and RN black hole solutions}
\label{sec_RNclas}

In this subsection the key features such as line element, divergent behavior and 
location of the horizons
of the classical (A)dS and RN black hole solutions
will be listed.
The line element of both of those black hole solutions takes the form 
\be\label{lineele}
ds^2= -f(r) \, dt^2+ f(r)^{-1} \, dr^2 + r^2 d\Omega_2^2\quad,
\ee
where $d\Omega_2^2$ is the volume element of the two-sphere.

For the case of the (A)dS solution, the function $f(r)$ take the form
\be\label{frfct}
f(r)|_{(A)dS}= 1 - \frac{2 G_0 M_0}{r} - \frac{1}{3} \, \Lambda_0 \, r^2 \, . 
\ee
Here $G_0$ and $\Lambda_0$ denote the classical Newton constant and 
the classical cosmological constant. 
The integration constant $M_0$ is the classical mass of the black hole. The sign of $\Lambda_0$ describes,
a Schwarzschild-AdS ($\Lambda_0 < 0$), Schwarzschild $(\Lambda_0 = 0)$,
or a Schwarzschild-dS ($\Lambda_0 > 0$) black hole. 
The previously mentioned space-like singularity at $r = 0$
can be seen for the (A)dS solution
 by computing the invariant square of the Riemann tensor
\be\label{singClass}
R_{\m\n\rho\sigma} R^{\m\n\rho\sigma}|_{(A)dS} = \frac{48 G_0^2 M_0^2}{r^6} + \frac{8 \Lambda_0^2}{3} \, .
\ee
This singularity is hidden behind an event horizon.
Horizons are found as zeros of the function $f(r)$, which due to the cubic nature of the function,
allows for three solutions
\be\label{AdSroot}
r_{ 0}|_{(A)dS} = -U^{-1/3} - \frac{1}{\Lambda_0} U^{1/3} \, , \quad
\ee
and
\be\label{dSroot}
r_{\pm}|_{(A)dS} = \frac{1}{2} \left( 1 \pm i \sqrt{3} \right) U^{-1/3}  + \frac{1 \mp i \sqrt{3}}{2 \Lambda_0} U^{1/3} \, . 
\ee
Where 
\be
U = 3 G_0 M_0 \Lambda_0^2 + \sqrt{ 9 G_0^2 M_0^2 \Lambda_0^4 -\Lambda_0^3}\quad,
\ee
was defined as the typical scale of the solution.
If the value of those roots is real and positive, they correspond to a physical horizon.
For the case of $\Lambda_0 \le 0$ there is only a single horizon, given by (\ref{AdSroot}).
For the case of $\Lambda_0 >0$ and $M_0>0$ the solution has two physical horizons
given by (\ref{dSroot}).\\

For the case of the classical RN solution, the function $f(r)$ in the line element (\ref{lineele})
takes the form
\be\label{frRN0}
f(r)|_{RN}=1-\frac{2G_{0}M_{0}}{r} + \frac{4\pi G_{0}Q_{0}^{2}}{r^{2}{e_{0}}^{2}}\quad,
\ee
where the constant of integration $Q_0$ is the classical electrical charge of the RN black hole
and $1/e_0^2$ is the electromagnetic coupling constant.
The classical solution for the electromagnetic stress energy tensor is
\be\label{Fmn0}
F_{tr}=-F_{rt}=\frac{Q_0}{r^2}\quad.
\ee
The leading singular behavior of the classical RN solution at $r=0$ 
can be seen by computing the invariant square of the Riemann tensor
\be
R_{\alpha \beta \gamma \delta}R^{\alpha \beta \gamma \delta}|_{RN}
= 8^2\frac{14 G_0^2 \pi^2 Q_0^4}{e_0^4 r^8}\quad.
\ee
Again, this singular behavior can be shielded by a horizon which can be found
by solving the condition of vanishing~(\ref{frRN0}) 
\be\label{RNhor0}
r_{\pm}|_{RN}= G_{0}M_{0} \pm \sqrt{{G_{0}} ^{2}{M_{0}}^{2} -\frac{4\pi {Q_{0}}^{2} }{{e_{0}}^{2} }}\quad.
\ee
One observes that those two horizons ($\pm$) become degenerate if the square root on the right hand side vanishes.
Even more, beyond this point the square root turns negative and no physical horizon is present in the solution, which
is undesired since it would lead to a unshielded ``naked'' singularity.
Thus, in order to not get in trouble due to the appearance of a naked singularity
one demands a minimal mass for the classical RN black hole.
\be\label{CosCens0}
M_0\ge 2 \sqrt{\pi}\frac{Q_0}{e_0 G_0}\quad.
\ee
This reasoning is known as the
 ``cosmic censorship'' argument \cite{Penrose:CS}.

\subsection{Scale dependent couplings and scale setting}
\label{sec_scaleset}

This subsection summarizes the equations of motion for the scale dependent
Einstein-Hilbert-Maxwell system.
The notation follows closely \cite{Koch:2014joa},
where also a more detailed description of this system and the proof of its self-consistency
can be found.
The three scale dependent couplings of the scale dependent
Einstein-Hilbert-Maxwell system are, the gravitational coupling
$G_k$, the cosmological coupling $\Lambda_k$, and the electromagnetic coupling
$1/e_k$.
Further, the system has three types of independent fields, which are
the metric field $g_{\mu \nu}(x)$, the electromagnetic four potential $A_\mu(x)$, and the
scale field $k(x)$.
The equations of motion for the metric field $g_{\mu \nu}(x)$ are
\begin{eqnarray}
 G_{\mu\nu}=-g_{\mu\nu}\Lambda_{k}-
\Delta t_{\mu \nu}+8 \pi \frac{G_k}{e_k^2} T_{\mu \nu}\quad,
\label{eomg}
\end{eqnarray}
where the possible coordinate dependence of $G_k$
induces an additional contribution to the stress-energy tensor
\cite{Reuter:2003ca}
\be
\Delta t_{\mu \nu}=G_{
k}\left(g_{\mu\nu}\Box-\nabla_\mu\nabla_\nu\right)\frac{1}{G_{k}}\quad.
\ee
The stress-energy tensor for the electromagnetic part is given by  
\be
T_{\mu \nu}=F_\nu^{\;\alpha} F_{\mu \alpha}- \frac{1}{4}g_{\mu \nu} F_{\alpha \beta}F^{\alpha \beta}\quad,
\ee
where $F_{\mu \nu}=D_\mu A_\nu- D_\nu A_\mu$ is the
antisymmetric electromagnetic field strength tensor.
The equations of motion for the four potential $A_\mu(x)$ are
\be\label{eomA}
D_\mu\left( \frac{1}{e_k^2 }F^{\mu \nu}\right)=0 \quad.
\ee
Finally the equations of motion for the scale-field $k(x)$ are given by
\be\label{eomk}
\left[R\nabla_\mu \left(\frac{1}{G_k}\right)-
2\nabla_\mu\left(\frac{\Lambda_k}{G_k}\right)
-F_{\alpha \beta}F^{\alpha \beta} \nabla_\mu \left( \frac{4 \pi}{ e_k^2}\right)\right]\cdot (\partial^\mu k)
=0 \quad.
\ee
The above equations of motion are complemented
by the relations corresponding to global symmetries of the system.
For the case of coordinate transformations one has
\be\label{diffeo}
\nabla^\mu G_{\mu \nu}=0
\ee
and for the internal $U(1)$ transformations
the corresponding relations are
\be\label{MaxwHom}
\nabla_{[ \mu} F_{\alpha \beta ]}=0\quad.
\ee
Please note that one has to work with (\ref{MaxwHom}) and not with
$\nabla_{[ \mu}e^{-2} F_{\alpha \beta ]}=0$ \cite{Koch:2014joa}.
In the following sections, two special black hole solutions for this system will be presented and discussed.
First, in section \ref{sec_EHsol}
a solution for the system (\ref{eomg}-\ref{eomk}) will be presented, where the electromagnetic coupling 
is omitted ($1/e_k^2=0$).
Then, in section \ref{sec_RNsol}, a solution is found for the case of finite electromagnetic coupling
and vanishing cosmological coupling ($\Lambda_k =0$).

\section{Black hole solution for the Einstein-Hilbert case}
\label{sec_EHsol}

In the Einstein Hilbert truncation one neglects
the electromagnetic contribution to the action of the system (\ref{eomg}-\ref{eomk}) leading
to simplified equations of motion for the metric field $g_{\mu \nu}$
\begin{equation}\label{eomgEH}
G_{\mu\nu}=-g_{\mu\nu}\Lambda _{k}-\Delta t_{\mu\nu} \quad,
\end{equation}
and simplified equations of motion for the scale field $k(x)$
\cite{Koch:2010nn}
\be\label{condi}
R\nabla_\mu \left(\frac{1}{G_k}\right)-
2\nabla_\mu\left(\frac{\Lambda_k}{G_k}\right)=0\quad.
\ee
The most general line element consistent with spherical symmetry is
\be\label{lineSph}
ds^2= -f(r) \, dt^2+ h(r) \, dr^2 + r^2 d\Omega_2^2\quad.
\ee
One notes that for this symmetry, the system (\ref{eomgEH}, \ref{condi}) has sufficient
independent equations, in order to solve for the three 
$r$-dependent functions $f(r)$, $h(r)$, and $k(r)$.
This is however, assuming that the functional form of the scale dependent
couplings $G_k$, and $\Lambda_k$ is known, for example from
a background independent integration of the functional renormalization group \cite{Weinberg:1979,Wetterich:1992yh,Dou:1997fg,Souma:1999at,Reuter:2001ag,Fischer:2006fz,Percacci:2007sz,Litim:2008tt}.

Since the aim is to gain some information on scale dependent black holes, 
independent of the details of the derivation and integration 
of the renormalization group or the particular approach to quantum gravity, we will use the following reasoning:\\
Even if one does not know the functional form of $G_k$ and $\Lambda_{k}$,
one knows that both couplings will inherit some $r$-dependence from $k(r)$ and therefore 
one might treat them
as two independent fields $G(r)$ and $\Lambda(r)$.
Thus, one has encoded the ignorance (or ambiguity) 
on the scale dependent couplings in an additional field variable
($G(r)$ and $\Lambda(r)$ instead of $k(r)$).
Of course, now the system (\ref{eomgEH}, \ref{condi}) with three independent equations
is in any case insufficient to solve for the four $r$-dependent fields $f(r),\, h(r),\;G(r)$, and $\Lambda(r)$
in full generality.
In order to reduce again the number free fields one has to impose some condition on those functions.
In the presented study we will restrict our search to solutions that have
only ``standard'' event horizons.  By this we mean that on the one hand
the signature of a ($t,r$) line will change from minus to plus or vice versa when passing 
an event horizon (zero of $f(r)$), which suggests that either $f(r)\sim h(r)$ or $f(r)\sim 1/h(r)$.
On the other hand this means that we demand that the radial part of the line element
diverges, when the time part of the line element vanishes.
Those conditions are implemented straight forwardly by imposing
\be\label{restfh}
f(r)\sim \frac{1}{h(r)}\quad.
\ee
This choice is commonly referred to as ``Schwarzschild ansatz''.
With the external restriction (\ref{restfh}) the number of fields is thus reduced 
to three:  $f(r),\;G(r)$, and $\Lambda(r)$, which fits the number of independent equations in 
the system (\ref{eomgEH}, \ref{condi}).

\subsection{Finding the solution}
\label{sec_findAdS}

Based on the condition (\ref{restfh}), the ansatz for the line element in the Einstein-Hilbert case will be
\begin{equation}\label{ansatz}
ds^{2}=-f(r)c_t^2 dt^{2}+\frac{1}{f(r)}dr^{2}+ r^2 d\Omega_2^2\quad,
\end{equation}
where the constant $c_t$ implements explicitly the time-reparametrization invariance of the system.
The equations (\ref{eomgEH}, \ref{condi})  have already been solved
\cite{Contreras:2013hua,Koch:2013rwa,Rodriguez:Ubu} by using the ansatz (\ref{ansatz}) for $c_t=1$.
However, in the parametrization found in \cite{Contreras:2013hua,Koch:2013rwa,Rodriguez:Ubu}, the
physical meaning of the integration constants and their relation to the classical
(A)dS-Schwarzschild metric remained unclear.

Here, a new parametrization (using the labels $G_0, \Lambda_0, M_0$, and $\epsilon$, $c_t$, and $c_4$)
of the solution is presented, where those problems were solved.
The solutions for the three functions are
\begin{eqnarray}\label{GrEH}
G(r)&=& \frac{G_0}{\epsilon r+1} \\ \label{frEH}
f(r) &=&1+3G_0M_0 \epsilon -\frac{2G_0M_0}{r} -(1 +6 \epsilon G_0M_0)\epsilon r -\frac{\Lambda_0 r^{2}}{3} + r^{2}\epsilon^{2}(6\epsilon G_0M_0+1 )\ln \left(\frac{c_4(\epsilon r +1)}{r}\right)\\ \label{LrEH}
\Lambda (r)&=&\frac{-72 \epsilon ^{2}r (\epsilon r+1) \left(\epsilon r+\frac{1}{2}\right) \left(G_0M_0\epsilon +\frac{1}{6} \right) \ln \left(\frac{c_4(\epsilon r +1)}{r}\right)+4r^{3}\Lambda_0\epsilon ^{2} + \left(12\epsilon ^{3} +6\Lambda_0 \epsilon +72\epsilon ^{4}G_0M_0 \right) r^{2}   }{2r(\epsilon r+1)^{2}} \\ \nonumber
&& +\frac{\left(72 \epsilon ^{3}G_0M_0 +11\epsilon ^{2} +2\Lambda_0 \right) r +6\epsilon ^{2}G_0M_0 }{2r(\epsilon r+1)^{2}}\quad.
\end{eqnarray}
The new and intuitive feature of this non-trivial choice of constants of integration is given by the fact
that it was possible to isolate a combination of the original constants of integration such that
the classical (A)dS solution can be recovered by sending a single constant ($\epsilon$) to zero.
 For instance, the scale dependent Newton coupling reduces to the classical Newton constant
\begin{equation}\label{limG}
\lim _{\epsilon \rightarrow 0} G(r) = G_0 \quad
\end{equation}
and the scale dependent cosmological coupling reduces to the cosmological constant
\begin{equation}\label{limL}
\lim _{\epsilon \rightarrow 0} \Lambda (r) = \Lambda_0 \quad.
\end{equation}
Finally, the labeling of the constant $M_0$ is justified
by taking the same limit for the metric component
\begin{equation}\label{limf}
\lim _{\epsilon \rightarrow 0} f(r) = - \frac{\Lambda_0 r^{2}}{ 3} - \frac{2G_0M_0}{r} +1 \quad,
\end{equation}
where the classical solution (\ref{frfct}) and the mass $M_0$ is recovered.
Thus, the limits  (\ref{limG}, \ref{limL}, and \ref{limf})  justify the choice of constants of integration as
$G_0, \Lambda_0, M_0$, and $\epsilon$ in contrast to the parametrizations found in \cite{Contreras:2013hua,Koch:2013rwa,Rodriguez:Ubu}.
Please note that the problem with finding this new parametrization was that
the limit $\epsilon \rightarrow 0$ corresponds to sending two constants of the parametrization \cite{Contreras:2013hua,Koch:2013rwa,Rodriguez:Ubu}
simultaneously and at a specific rate to infinity.

\subsection{Asymptotic space-times}
\label{sec_AdSasympt}

The asymptotic behavior of this solution for small scales ($r \rightarrow 0$) is closely linked
to the singularity at the origin. This singularity can be most clearly studied by evaluating
geometrical invariants.
For example the Ricci scalar for the solution is
\be
R= -\frac{6 G_0 M_0 \epsilon}{r^2}+\frac{6 \epsilon+ 36 G_0 M_0 \epsilon^2}{r}
+ 7 \epsilon^2+42 G_0 M_0 \epsilon^3 + 4 \Lambda_0-12 \epsilon^2(1+6 G_0 M_0 \epsilon)\log \left( \frac{c_4}{r}\right)
+{\mathcal{O}}(r)\quad.
\ee
One observes that the classical limit of $4 \Lambda_0$ is modified
by a new quadratic divergence of this quantity which is proportional to $\epsilon$.
An other invariant quantity that is frequently studied is the higher curvature scalar
\be
R_{\mu \nu \alpha \beta}R^{\mu \nu \alpha \beta}
=\frac{48 G_0^2 M_0^2}{r^6}-
\frac{48 G_0^2 M_0^2 \epsilon}{r^5}+{\mathcal{O}}(r^{-4})\quad.
\ee
For this invariant one observes that to leading order the singular behavior of this
quantity is the same as in the classical case (\ref{singClass})
and that modifications due to $\epsilon$ only appear at subleading
orders in $1/r$.
From those two examples one can already conclude that the elimination of this radial singularity
is not possible for the given solution unless one returns to trivial configurations - say
with vanishing $M_0$ and vanishing $\epsilon$.

The other regime of asymptotic behavior can be studied in a large large radius expansion $r \rightarrow \infty$.
In order to get a feeling for this behavior
it is instructive to plot the radial function $f(r)$ for  varying values of $\epsilon$.
 \begin{figure}[hbt]
   \centering
\includegraphics[width=10cm]{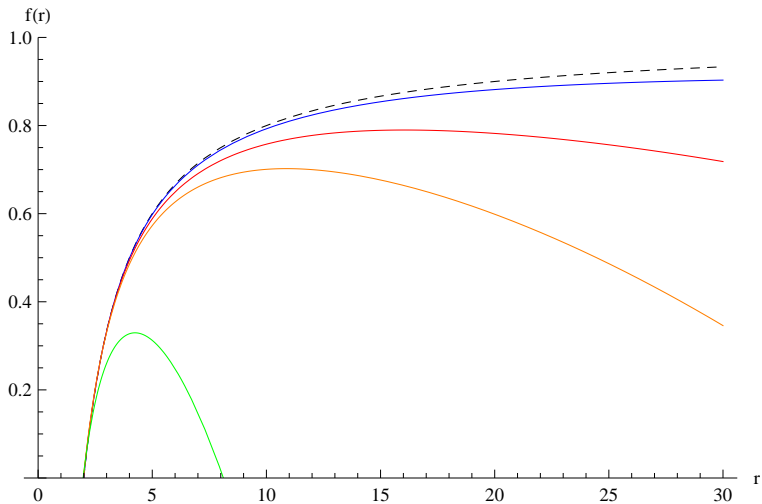}
  \caption{Radial function $f(r)$ for $G_0=1$, $M_0=1$, $\Lambda_0=0$, and $\epsilon = \{0,0.001, 0.005, 0.01, 0.05\}$}
\label{fig:frL0}
\end{figure}
In figure \ref{fig:frL0}
one observes that even for the case $\Lambda_0=0$, the non-classical scale dependence
$\epsilon \neq 0$ can mimic the effect of a 
cosmological constant by generating an asymptotic (Anti)-de-Sitter space-time.

By studying the large $r$ behavior of the metric function (\ref{frEH})
 one finds
 \be
 f(r)= - r^2 \frac{\tilde \Lambda}{3}+ {\mathcal{O}}(r)\quad,
 \ee
where the effective cosmological constant $\tilde \Lambda$
is actually a shift of the classical value
\be\label{tildeL}
\tilde \Lambda = \Lambda_0 - 3 \epsilon^{2}(6\epsilon G_0M_0+1 )\ln \left(c_4 \epsilon \right)\quad.
\ee
When it is referred to $\tilde \Lambda$ as ``effective cosmological constant'',
this is done in the sense that for any measurement, say by observing trajectories,
the result would be determined by the
form of the metric function rather than by the function $\Lambda(r)$ in (\ref{LrEH}).
One observes that the $\epsilon$ induced shift in the cosmological constant $\tilde \Lambda - \Lambda_0$
 is determined by the sign of the logarithm in (\ref{frEH}).
In the asymptotic limit $r \rightarrow \infty$ one finds for example for
$\epsilon > -\frac{1}{6 G_0 M_0}$ that for
\bea\label{AdSrelations}
c_4 < \frac{1}{\epsilon} & \Rightarrow & (\tilde \Lambda -\Lambda_0)> 0 \quad,\\ \nonumber
 c_4 =\frac{1}{\epsilon} & \Rightarrow & (\tilde \Lambda -\Lambda_0)= 0\quad,\\ \nonumber
 c_4 > \frac{1}{\epsilon} & \Rightarrow & (\tilde \Lambda -\Lambda_0)< 0\quad.
\eea
For $\epsilon < -1/(6 G_0 M_0)$ the relations (\ref{AdSrelations}) get inverted.
A very interesting scenario turns out to be the case of $c_4 = \epsilon$,
where the effective cosmological constant agrees with the classical parameter $\tilde \Lambda = \Lambda_0$.
However, studying the asymptotics of the metric function (\ref{frEH}) is not the only way one might
try to extract a notion of an asymptotic cosmological constant.
For a comparison one can take the limit of large $r$ for the scale dependent quantity (\ref{LrEH})
which gives
\be\label{tildeL2}
\lim_{r \rightarrow \infty} \Lambda(r)= 2 \Lambda_0 -  6 \epsilon ^{2} (6 G_0 M_0 \epsilon+1) \ln \left(c_4 \epsilon \right) \quad,
\ee
which is different from the ``effective cosmological constant'' (\ref{tildeL}) extracted from the
metric solution.
This is however not concerning, since one can argue that the asymptotic form of a space-time
must be read from the metric and not from a function appearing in the equation of motion.
Based on this argument one sticks to (\ref{tildeL}) as the proper definition of $\tilde \Lambda$.
Still it is interesting to note that both possible notions of an ``effective cosmological constant''
(\ref{tildeL} and \ref{tildeL2}) vanish for the same choice of parameters
\bea
{\mbox{if}}
\;\Lambda_0 = 3\epsilon ^{2} (6\epsilon G_0M_0+1 )\ln \left(c_4 \epsilon \right)
&\Rightarrow & \lim_{r \rightarrow \infty} \Lambda(r)=\tilde \Lambda=0 \quad.
\eea
Further one observes that the $\epsilon$ dependence of both
notions vanishes for the particular choice $c_4=1/\epsilon$.

\subsection{Perturbative analysis for horizons and thermodynamics}
\label{sec_AdSthermo}

Since scale dependence of coupling constants is generally
assumed to be weak, it is reasonable to treat the dimensionful parameter $\epsilon$ as small
with respect to the other scales entering the problem such as $1/\sqrt{G_0}$, or $M_0$.
As it can be seen from the relations (\ref{limG}-\ref{limf}), this constant encodes the deviation
from the classical solution (\ref{frfct}) and therefore its absolute value is also experimentally expected to be very small in comparison with other integration constant with dimensions of energy.
In principle $\epsilon$ could take positive or negative values.
However, the following short discussion will show that only small 
positive values give physically viable (real) solutions at the outside of the event horizon:

From the solution (\ref{frEH}) one sees that the argument of logarithm in the metric function could become negative for $\epsilon < 0$ and $c_4>0$,
at very large values of $r$.
Compensating this by making $c_4<0$ is also not possible since in this case the logarithm can become negative for somewhat smaller radii $r_S < r<1/|\epsilon|$,
where $r_S$ is the radius of the horizon which for small $|\epsilon |$ can be approximated by the classical Schwarzschild radius.
Thus, the parameter $\epsilon$ has to be positive and small right from the start.

In this context it is instructive to Taylor expand in this small parameter
to see the leading corrections due to the scale dependence of the couplings
 \begin{eqnarray}
 G(r)&=& G_0 - \epsilon \cdot G_0 r+{\mathcal{O}}(\epsilon^2)  \\ \label{frapproxEH}
f(r) &=&1-\frac{2 G_0 M_0}{r}-\frac{r^2 \Lambda_0}{3}+\epsilon \cdot (3 G_0 M_0- r) +{\mathcal{O}}(\epsilon^2)\\
\Lambda (r)&=&\Lambda_0 +\epsilon \cdot \Lambda_0 r +{\mathcal{O}}(\epsilon^2)\quad.
 \end{eqnarray}
From equation (\ref{frapproxEH}) one sees now more clearly why previous attempts
to obtain a meaningful physical parametrization failed.
The problem
was that it was assumed that for vanishing parameter $M_0$
the flat (A)dS solution $f(r)=1 -\frac{r^2 \Lambda_0}{3} $ would be recovered.
However, looking at (\ref{frapproxEH}) one sees that this
is actually not possible without completely returning to the classical solution ($\epsilon=0$).
Apparently the deviations from the classical space-time metric in (\ref{frapproxEH}),
could be used in a phenomenological context in order to constrain the value of the
supposedly very small parameter $\epsilon$.
Such a study is however beyond the scope of this work.
The perturbative analysis is however useful for a first understanding
of the leading effects on the black hole horizons and the corresponding
thermodynamics.

\begin{figure}[hbt]
   \centering
\includegraphics[width=10cm]{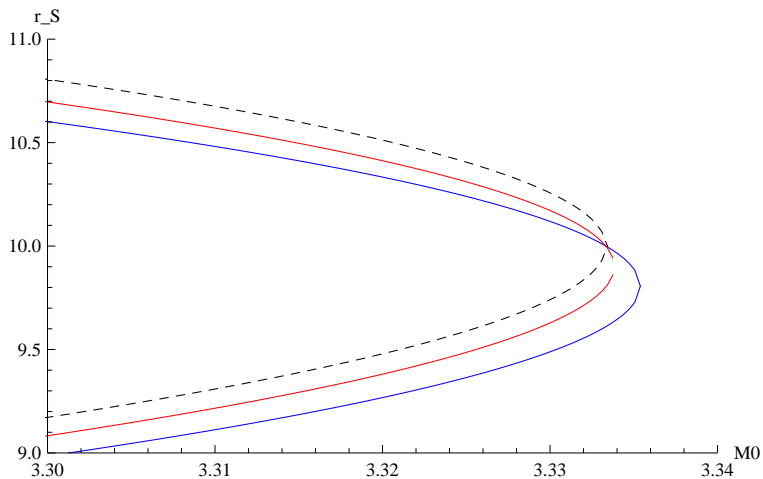}
  \caption{\label{figHorPosEH}
  Mass dependence of the black hole horizons $r_{\pm}$
  as a function of the mass parameter $x=M_c/M_0$, for
  $G_0=1$ and  $\Lambda_0=0.01$.
  The black dashed curve is for $\epsilon=0$ the red curve is for $\epsilon=0.002$
  and the blue curve is for $\epsilon=0.004$.
  }
\end{figure}

To first order in $\epsilon$, the horizons are defined by the zeros of (\ref{frapproxEH}).
For $\Lambda_0 > 0$ and for  $0 < M_0 < \frac{1}{3 G_0 \sqrt{\Lambda_0}}$ the two relevant real horizons are found to be
\be \label{rPN}
r_{\pm}=\frac{1\pm i \sqrt{3}}{2 P^{1/3}}+\frac{(1\mp i \sqrt{3})P^{1/3}}{2 \Lambda_0}+
\epsilon \left(
\frac{1\mp i \sqrt{3}}{4 P^{2/3}}+\frac{(1\pm i \sqrt{3})(6 G_0 M_0 P-\Lambda_0)}{4 P^{4/3}}
-\frac{1}{\Lambda_0}
\right)\quad,
\ee
where $r_{+}$ is the outer (cosmological) horizon and $r_{-}$ is the inner
Schwarzschild horizon. $P$ is given by
\be
P = 3 G_0 M_0 \Lambda_0^2 \left( 1 + \sqrt{(1- \frac{1}{9 G_0^2 M_0^2 \Lambda_0})}\right)\quad.
\ee
In the classical case ($\epsilon=0$) those two horizons become degenerate for the critical mass
\be\label{MNari}
M_{0,crit}= \frac{1}{3 G_0 \sqrt{\Lambda_0}}\quad,
\ee
which corresponds to the Nariai black hole \cite{Nariai}, 
which is the maximal allowed black hole mass before the
appearance of a naked singularity.
For non-vanishing $\epsilon$ this critical mass value and the corresponding
black hole radius get slightly shifted. A comparison of the horizon structure
as a function of the mass parameter $M_0$ is shown in figure \ref{figHorPosEH}.

One observes that for a given $M_0$ both horizons get shifted by positive $\epsilon$
towards smaller radii. Further, one sees that the same holds true for the
radius of the critical Nariai black hole and that
the critical mass parameter $M_0$ which is the value where the cosmological and the
inner horizon merge $r_+=r_-$ gets slightly increased with respect to the classical value (\ref{MNari}).

\begin{figure}[t]
  \centering
\subfloat[\hspace{7cm}.
$.\quad$Mass range $M_0=0 \dots 4$ ]{\label{figTHvsMEH1}
\includegraphics[width=0.48\textwidth]{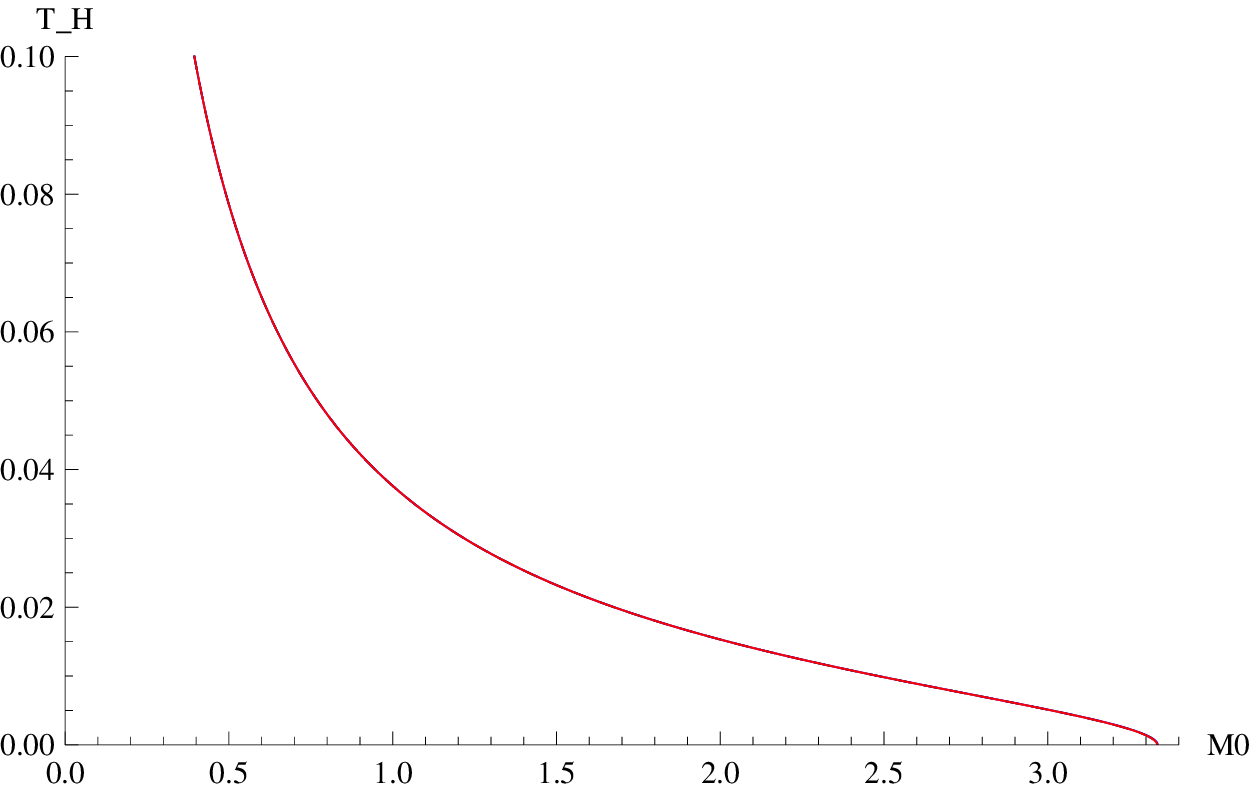}}
\hspace{0.1cm}
\subfloat[\hspace{7.0cm}.
$.\quad$Mass range $M_0=3.3 \dots 3.34$]{\label{figTHvsMEH2}
\includegraphics[width=0.48\textwidth]{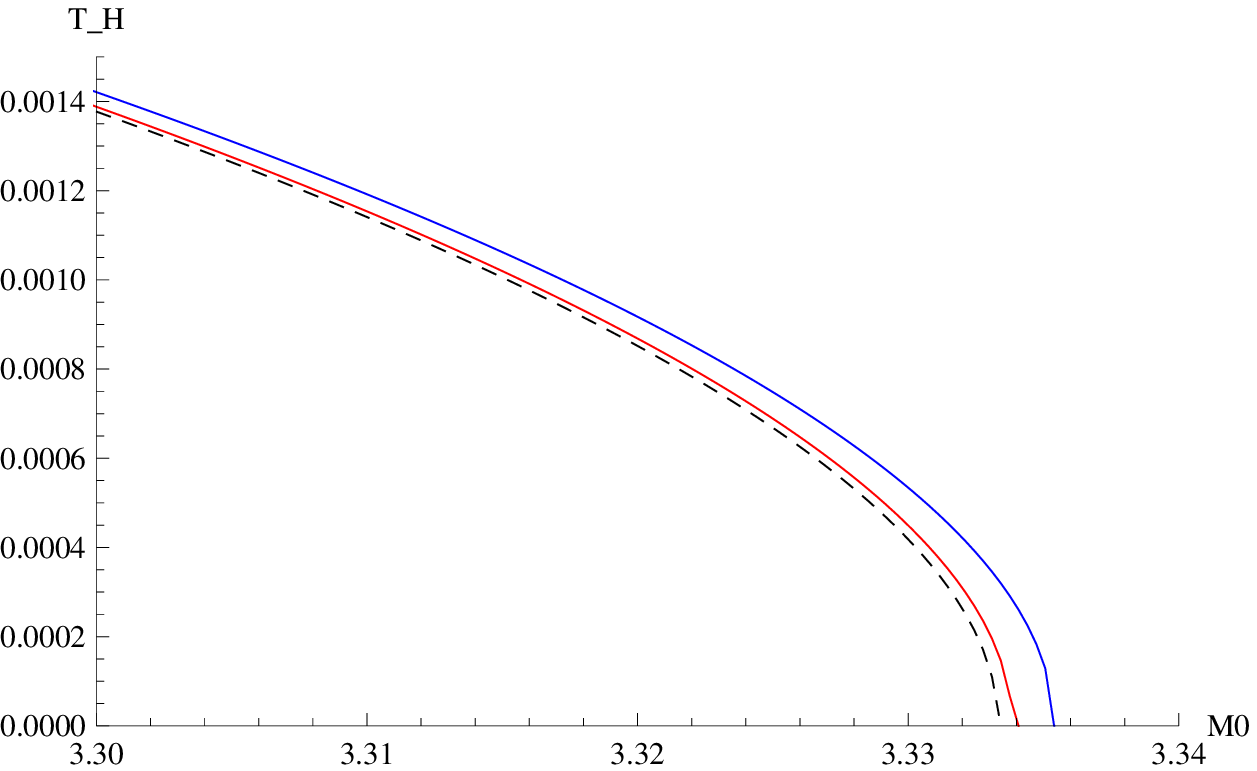}}
\caption{
  $M_0$ dependence of the temperature (\ref{tempEH})
  for  $G_0=1$ and  $\Lambda_0=0.01$.
  The black dashed curve  corresponds to $\epsilon=0$, the red curve is for $\epsilon=0.002$
  and the blue curve is for $\epsilon=0.004$.
}
\label{figTHEH}
\end{figure}

Given the horizon structure and the functional form of (\ref{frapproxEH})
one can calculate the temperature of corresponding black hole.
At the inner horizon this temperature is given by
\be\label{tempEH}
T_{r_-}=\frac{1}{4 \pi} \left.\frac{df}{dr}\right|_{r_-}=
\frac{2 G_0 M_0}{r_-^2}-\epsilon-\frac{2 r_- \Lambda_0 }{3}+
{\mathcal{O}}(\epsilon^2)\quad.
\ee
In figure \ref{figTHEH} this temperature is shown as a function of the mass parameter $M_0$.
One observes that for vast range of parameters the modified temperature
is indistinguishable from the classical value, and that only for the largest masses, close
to the $M_c$, a slight splitting of the curves occurs.
This splitting shows a slightly increased temperature for increasing $\epsilon$ values.
Thus, in this region, the shift in the horizon radius $r_-$ overcompensates the
negative direct contribution of $\epsilon$ to the temperature in equation (\ref{tempEH}).
Therefore it takes slightly higher values of $M_0$ in order to reach
zero temperature, and thus the critical black hole state,
as it can also be seen from the figures \ref{figHorPosEH} and \ref{figTHEH}b .

\newpage

\section{Black hole solution for the Einstein-Maxwell case}
\label{sec_RNsol}

In this section, a black hole solution for the Einstein-Maxwell case will be constructed
without taking into account the cosmological term ($\Lambda_k=0$).
In this case the equations of motion for the metric field (\ref{eomg})
simplify to
\begin{eqnarray}
 G_{\mu\nu}=-\Delta t_{\mu \nu}+8 \pi  \frac{ G_k}{{e_{k}} ^{2}} T_{\mu \nu}\quad,
\label{eomgEM}
\end{eqnarray}
while equations of motion (\ref{eomA}) for the $U(1)$ gauge field remain unchanged
\be\label{eomA2}
D_\mu\left( \frac{1}{e_k^2 }F^{\mu \nu}\right)=0 \quad.
\ee
The equations of motion for the scale-field $k$ (\ref{eomk})
simplify to
\be\label{eomk2}
\left[R\nabla_\mu \left(\frac{1}{G_k}\right)
-\nabla_\mu \left( \frac{4 \pi}{ e_k^2}\right)F_{\alpha \beta}F^{\alpha \beta}\right]\cdot (\partial^\mu k)
=0 \quad.
\ee
The invariance equations (\ref{diffeo}) and (\ref{MaxwHom}) remain unchanged.

\subsection{Finding the solution}
\label{sec_RNfind}

When searching a solution of the above equations we proceed by imposing spherical symmetry.
For spherical symmetry, the most general line element is again (\ref{lineSph}).
Assuming electric and not magnetic charge,
this symmetry requirement reduces the degrees of freedom of the electromagnetic
stress-energy tensor to
\be\label{Frt}
F_{tr}=-F_{rt}= q(r)   \quad.
\ee 
Under those assumptions and  for a given scale dependence $G_k$ and $1/e_k^2$ the system (\ref{eomgEM}-\ref{eomk2}) contains
four unknown functions $f(r)$, $h(r)$, $q(r)$, and $k(r)$.

Now, the reasoning of section (\ref{sec_EHsol}) will be repeated and
the ignorance (at least model dependence) of the coupling flow will be encoded in trading
the radial scale dependence $k=k(r)$ for radial coupling dependence
$k(r)\rightarrow G(r),\; 1/e^2(r)$.
The increase in unknown functions will be compensated by imposing  
the ``standard black hole'' condition (\ref{restfh}).
Due to this, the system (\ref{eomgEM}-\ref{eomk2}) will have to be solved for the
four functions $f(r)$, $q(r)$, $G(r)$, and $1/e^2(r)$.
The ansatz for the line element will be
\begin{equation}\label{ansatz2}
ds^{2}=-f(r)  dt^{2}+\frac{1}{f(r)}dr^{2}+ r^2 d\Omega_2^2\quad.
\end{equation}
Note that here, in contrast to (\ref{ansatz}), the constant $c_t$
is set to one, since due to the $F_{\mu \nu}$ contribution 
one can not expect to have time-rescaling invariance of the solution.

A good starting point for solving the system is to observe that $f(r)$
actually decouples from the equations for the radial electric field
\be\label{MaxRad}
\frac{1}{r^{2}}\frac{\partial}{\partial r}\left(r^2 \frac {q(r)}{e^2(r)}\right)= 0\quad.
\ee
This establishes a first relation between $e(r)$ and $q(r)$.
With this, the remaining equations of motion (\ref{eomgEM} and \ref{eomk2}) are solved by
\begin{eqnarray}\label{solRN}
G(r)&=&\frac{G_{0}}{\epsilon r +1}\\ \nonumber
f(r)&=&\frac{r^{4}\epsilon ^{2}{e_{0} }^{2} 
+ 4\epsilon r^{3}{e_{0} }^{2}+ 4
(1-G_{0}M_{0}\epsilon ){e_{0} }^{2}r^{2}-8r G_{0}M_{0}{e_{0} }^{2}+16\pi G_{0}{Q_{0}}^{2}}{4r^{2}(\epsilon r +1)^{2} {e_{0} }^{2}} \\  \nonumber
e^2(r)&=&\frac{\left(r^{6}\epsilon ^{4}{e_{0} }^{2} + 3r^{5}\epsilon ^{3}{e_{0} }^{2} + \left(3r^{4}{e_{0} }^{2}-4r^{3}{e_{0} }^{2} G_{0}M_{0}+48r^{2} \pi G_{0}{Q_{0}}^{2} \right)\epsilon ^{2}+48\epsilon r \pi G_{0}{Q_{0}}^{2} +16\pi G_{0} {Q_{0}}^{2} \right) {e_{0} }^{2} \pi}{ {Q_{0}}^{2} G_{0} (\epsilon r +1)^{3}} \\ \nonumber
q(r)&=& \frac{Q_{0}}{4\pi e_0^{2}}\frac{e^2(r)}{r^2}\quad.
\end{eqnarray}
Having learned the lesson from section \ref{sec_EHsol} on the subtleties of choosing the constants of integration,
the five constants of integration were chosen such that, for the case that the fifth constant vanishes,
the four other constants correspond to constants in the classical solution of $f(r)|_{RN}$.
This allows to interpret this fifth constant,
which again will be labeled $\epsilon$, as deviation parameter which introduces corrections to the
classical solution due to scale dependence.
One confirms for Newtons coupling
\begin{equation}\label{limGRN}
\lim _{\epsilon \rightarrow 0} G(r) = G_0 \quad,
\end{equation}
and for the metric function
\be
\lim _{\epsilon \rightarrow 0} f(r)=1-\frac{2G_{0}M_{0}}{r} + \frac{4\pi G_{0}Q_{0}^{2}}{r^{2}{e_{0}}^{2}} \quad,
\ee
which reproduces the classical solution (\ref{frRN0}).
Similarly one finds for the scale dependent electrical coupling
\be
\lim _{\epsilon \rightarrow 0} e^2(r)=(4\pi)^2 e_0^2 
\ee
and the electrical field strength
\be
\lim _{\epsilon \rightarrow 0} q (r)= \frac{4\pi Q _{0} }{r^{2}}\quad.
\ee
One observes that this is also in agreement with the classical
result. The factor $(4 \pi)$, that appears as different normalization of $q(r)$
is a convention which turns out to cancel for the ratios $q^2(r)/e^2(r)$
that enter the equations of motion (\ref{eomgEM} and \ref{eomk2}). 
This convention is further justified when calculating the actual charge of the solution (\ref{solRN}).
Taking again $\epsilon$ smaller than any dimensionfull scale of the system, one can 
expand in this parameter. In this expansion the lowest order corrections to the classical solution are
\begin{eqnarray}
G(r) &=&G_{0}\,\,-\epsilon G_0 r  +{\mathcal{O}}\left(\epsilon^2 \right)\quad,\\ \nonumber
f(r)&=&1-\frac{2G_{0}M_{0}}{r} + \frac{4\pi G_{0}Q_{0}^{2}}{r^{2}{e_{0}}^{2}}
\,\,+\epsilon \left( 3 G_0 M_0-\frac{8 G_0 \pi Q_0^2}{e_0^2 r}-r\right)
+{\mathcal{O}}\left(\epsilon^2 \right)\quad,\\  \nonumber
e^2(r)&=&e_0^2 16\pi ^{2}
\,\,-\epsilon^2 \frac{e_0^4 \pi (4 G_0 M_0 r^3-3 r^4)}{G_0 Q_0^2}+{\mathcal{O}}\left(\epsilon^3 \right)\quad,\\ \nonumber
q (r)&=& \frac{Q _{0} 4\pi}{r^{2}}
\,\,-\epsilon^2 \frac{e_0^2 (4 G_0 M_0 r-3 r^2)}{4 G_0 Q_0}
+{\mathcal{O}}\left(\epsilon^3 \right)\quad.
\end{eqnarray}
Fortunately, due to the purely polynomial form of the solution,
most of the following results can be discussed with the complete solution (\ref{solRN}),
without the necessity to use the above expansion.

\subsection{Asymptotic behavior of the solution}
\label{sec_RNasym}

The asymptotic behavior of this generalized Reissner-Nordstr\"om solution for $r\rightarrow 0$
can be studied by evaluating curvature invariants in this limit.
For example, already the Ricci scalar shows a quadratic divergence for small radii
\be\label{RNRicS}
R=-6G_{0}\frac{M_{0}\epsilon + 4\pi Q_{0}^{2}\epsilon ^{2} {e_{0}}^{-2} }{r^{2}}+{\mathcal{O}}\left(1/r \right)\quad.
\ee
This is in contrast to the classical solution where the Ricci scalar vanishes
in this limit.

Still the generalized solution incorporates the classical result,
since in the classical limit $\epsilon \rightarrow 0$, the right hand side of (\ref{RNRicS}) vanishes accordingly.
The invariant contraction of two Riemann tensors is also divergent in this
limit, but for this invariant, the leading divergence agrees with the classical behavior
\be\label{RNsing2}
R_{\alpha \beta \gamma \delta}R^{\alpha \beta \gamma \delta}
= 2^7 7\frac{G_0^2 \pi^2 Q_0^4}{e_0^4 r^8}+{\mathcal{O}}\left(1/r^7 \right)\quad.
\ee
This confirms that this solution is singular at the origin for the generalized solution,
even in the classical limit.

Taking the opposite limit (for $r \rightarrow \infty$), the asymptotic behavior in  brings a surprise since
the metric function approaches
\be\label{RNfrgrand}
f(r)= \frac{1}{4}+ \frac{1}{2 \epsilon r}+{\mathcal{O}}\left(1/r^2 \right)\quad.
\ee
This result does not resemble the classically expected value $1$, 
not even by approaching a posteriori $\epsilon \rightarrow 0$.
The supposed discrepancy can be explained by the fact all
dimensionless terms $\epsilon r$ are incompatible with first taking the limit of
large $r$ and than the limit of small $\epsilon$.
Clearly, if one takes the limit of $\epsilon \rightarrow 0$ first,
the classical result is recovered.
The asymptotic line element corresponding to (\ref{RNfrgrand}) is
\begin{equation}\label{monop1}
ds^{2}_{\infty}=-\frac{1}{4}dt^{2}+4dr^{2}+r^{2}d\theta^{2}+r^{2}\sin^{2} \theta d\phi ^{2} \quad.
\end{equation}
One can try to cast this in a more familiar form by introducing a 
rescaled time $\tau = \frac{1}{2} t$ and a rescaled radial coordinate $R=2 r$, giving the line element
\begin{equation}
ds^{2}_{\infty}=-d\tau ^{2}+dR^{2}+R^{2}\frac{1}{4}\left(d\theta^{2}+\sin^{2} \theta d\phi ^{2} \right)\quad.
\end{equation}
However, even though now the radial and temporal part of the line element take the familiar form, the angular part suffers a non-trivial scaling, which corresponds to a deficit solid angle.
For example, the area of a sphere with very large radius R is not $4\pi R^{2}$ but rather $\pi R^{2}$. 
The remaining factor of $1/4$ might be absorbed in a rescaling of the angles $\theta$ and $\phi$,
but this clearly also implies the mentioned deficit angle of the asymptotic geometry.

A very similar asymptotic behavior for very large radial distance is known from so called
global monopoles  \cite{Barriola:1989nn,Shi:2009nn}. Even though the above solution
is to our knowledge not present in the literature, its asymptotic behavior for large $r$
can be matched to the monopole in \cite{Barriola:1989nn} by identifying $\frac{3}{4} \frac{1}{8\pi G}$
with the monopole parameter $\eta ^{2}$, and $\frac{-1}{4G\epsilon}$ with the mass parameter of the monopole $\tilde M$.
However,  for the presented solution, 
negative values for $\epsilon$ do not allow a well defined classical limit and therefore the
asymptotic results discussed here do not apply for the global monopoles in \cite{Barriola:1989nn} and vice versa.

\begin{figure}[hbt]
   \centering
\includegraphics[width=10cm]{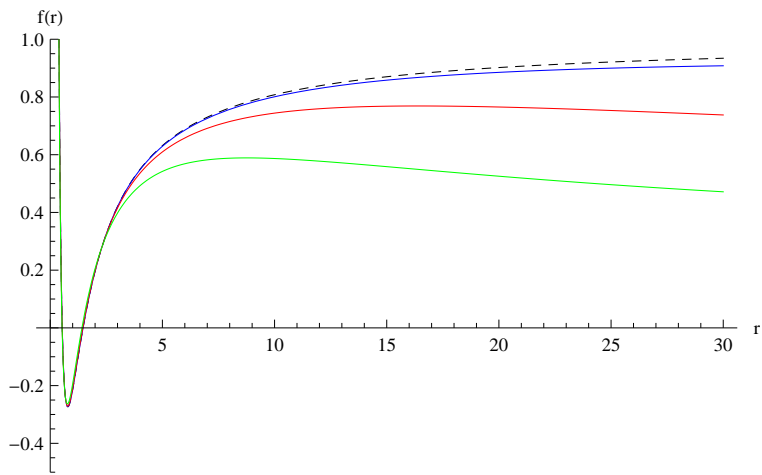}
  \caption{Radial function $f(r)$ for $G_0=1$, $M_0=1$, $e^2_0=1$, $Q_0=0.25$, and $\epsilon = \{0,0.001, 0.01, 0.05\}$,
  plotted in black-dashed, blue, red, and green respectively.}
\label{fig:frRN}
\end{figure}

In order to get an intuition on the radial dependence
of the radial function $f(r)$ and the corresponding asymptotic behavior one can also
refer to a graphical analysis, which is done
figure \ref{fig:frRN}.
One observes that corrections to the black dashed classical curve
become more and more prominent for large radii and increasing $\epsilon$.
One also observes that for small $\epsilon$, the metric function $f$ seems to approach
the classical value of one, before converging to the limit expressed in (\ref{RNfrgrand}), far outside of the shown region.

\subsection{Total charge}
\label{sec_RNcharge}

As written in the solution (\ref{solRN}), $Q_0$ is nothing more than an integration
parameter which inherits its interpretation as charge due to this classical limit $\epsilon \rightarrow 0$.
This does not determine the actual charge of the solution for values of $\epsilon \neq 0$.
Given the asymptotic deficit angle (\ref{monop1}), one might expect a corresponding
effect for the charge.
In curved space-time the actual charge corresponding to the Maxwell equation (\ref{eomA})
can be evaluated by the integral \cite{Carroll:2003st}
\begin{equation}\label{Qp0}
Q=\int _{\partial \Sigma} d^{2}z\sqrt{\gamma ^{S_{2}}} n_{\mu} \sigma _{\nu}\frac{F^{\mu \nu}}{e^2}\quad,
\end{equation}
where $n_{\mu}$ is the unitary vector associated to the time coordinate and
$\sigma_{\nu}$ is the unitary vector associated to the radial coordinate
\footnote{Please note that the calculation of the total black hole mass $M$ in terms of the classical mass
parameter $M_0$ is by far less straight forward and will be postponed to future studies}.
The integral over the surface $\partial \Sigma$ will be evaluated at radial infinity 
such that, according to the asymptotic metric (\ref{RNfrgrand}), the two dimensional surface element is
\be\label{Qp1}
d^{2}z\sqrt{ g^{S_{2}}}= d\theta d\phi r^{2} \sin(\theta)\quad,
\ee
with $\theta: 0 \dots \pi$ and $\phi: 0 \dots 2 \pi$.
From the same asymptotic metric  (\ref{RNfrgrand}) one reads that the properly normalized unitary vector time vector is
\be\label{Qp2}
n_\mu=(2,0,0,0)
\ee
and that the properly normalized radial vector is
\be\label{Qp3}
\sigma_\nu=(0,1/2,0,0)\quad.
\ee
The function associated to the electric field of the solution is 
\be\label{Qp4}
\frac{F^{tr}}{e^2}=\frac{Q_0}{4 \pi e_0^2}\frac{1}{r^2}\quad,
\ee
which is simply proportional to $1/r^2$, as it can
also be read directly from the relation (\ref{MaxRad}).
Putting (\ref{Qp0}-\ref{Qp4}) together 
One finds that all unusual factors of the generalized solution and the corresponding asymptotic metric  (\ref{RNfrgrand})
cancel out. The charge is
\be
Q=\frac{Q_0}{e_0^2}\quad,
\ee
which just resembles the classical value and has no $\epsilon$ dependence.

\subsection{Horizons, temperature, cosmic censorship}
\label{sec_RNtemp}

Important information on a black hole solution can be gained by studying its horizon structure
and the corresponding thermodynamic behavior. The possible horizons, which
correspond to zeros of $f(r)$ in (\ref{solRN}) are found to be
\begin{eqnarray}
r_{1} &=& \frac{-1-\sqrt{1+2 \epsilon G_{0} M_{0}-2 \epsilon \sqrt{{G_{0}}^{2} {M_{0}}^{2}-\frac{4  G_{0} \pi {Q_{0}}^{2}}{{e_{0}}^{2}}  }}}{\epsilon} \quad,\\
r_{2} &=& \frac{-1+\sqrt{1+2 \epsilon G_{0} M_{0}-2 \epsilon \sqrt{{G_{0}}^{2} {M_{0}}^{2}- \frac{4  G_{0} \pi {Q_{0}}^{2}}{{e_{0}}^{2}} }}}{\epsilon} \quad, \\
r_{3} &=& \frac{-1-\sqrt{1+2 \epsilon G_{0} M_{0}+2 \epsilon \sqrt{{G_{0}}^{2} {M_{0}}^{2}-\frac{4  G_{0} \pi {Q_{0}}^{2}}{{e_{0}}^{2}}  }}}{\epsilon} \quad,\\
r_{4} &=& \frac{-1+\sqrt{1+2 \epsilon G_{0} M_{0}+2 \epsilon \sqrt{{G_{0}}^{2} {M_{0}}^{2}-\frac{4  G_{0} \pi {Q_{0}}^{2}}{{e_{0}}^{2}}  }}}{\epsilon} \quad.
\end{eqnarray}
In the analysis of those horizons we will restrict to the case of $\epsilon> 0$, since it is this case that allows the transition to the classical values for 
$\epsilon \rightarrow 0$.
One sees that $r_1$ and $ r_3$ are always negative for positive $\epsilon$.
Thus, one defines the remaining horizons
\begin{equation}\label{RNrpm}
r_{\pm} =\frac{-1+\sqrt{1+2 \epsilon G_{0} M_{0} \pm 2 \epsilon \sqrt{{G_{0}}^{2} {M_{0}}^{2}-\frac{4  G_{0} \pi {Q_{0}}^{2}}{{e_{0}}^{2}} }}}{\epsilon} \quad.
\end{equation}
One confirms that those two horizons coincide with the classical horizons (\ref{RNhor0}) for $\epsilon \rightarrow 0^+$.
For vanishing charge ($Q_0 \rightarrow 0$), $r_-$ goes to zero and the remaining horizon is 
\be
r_{+}|_{Q_0=0}=\frac{-1+\sqrt{1+4\epsilon G_{0}M_{0}}}{\epsilon}\quad,
\ee
which gives in the classical limit the expected Schwarzschild value of $2G_{0}M_{0}$.

\begin{figure}[hbt]
   \centering
\includegraphics[width=10cm]{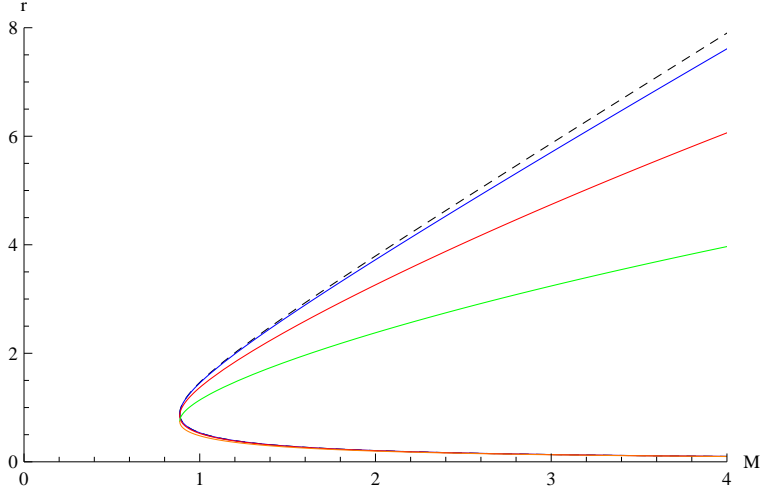}
  \caption{Horizon radius $r_\pm$ for $G_0=1$, $e^2_0=1$, $Q_0=0.25$, and $\epsilon = \{0,0.001, 0.1, 0.5\}$,
 plotted in black-dashed, blue, red, and green respectively.}
\label{fig:rMRN}
\end{figure}

As it was seen in the asymptotic limit of small radii (\ref{RNsing2}), the singularity at zero radius persists 
for the generalized solution. Therefore,
one still needs to invoke the cosmic censureship hypothesis in order to avoid the ``visibility'' of this naked singularity.
For the case of the present solution, this hypothesis can be addressed by studying the critical value 
for which the inner and outer horizon merge,
which is in the classical case given for (\ref{CosCens0}).
For the generalized solution the merging occurs when the inner square root in (\ref{RNrpm}) vanishes.
This is true for the critical value
\be\label{CosCens1}
M_0= 2 \sqrt{\pi}\frac{Q_0}{e_0 G_0}\quad,
\ee
which is exactly the classical ($\epsilon$ independent) value given in (\ref{CosCens0}).
There might be the possibility of merging horizons other than $r_\pm$,
but this possibility can be ignored since one already knows that $r_1, r_3 \le 0$,
which at most would allow merging horizons at the origin.
The behavior of the physical horizons (\ref{RNrpm}) and their merging
at the classical horizon value in is shown figure \ref{fig:rMRN}.
One observes nicely that the mass-value of the critical horizon is independent of the value of $\epsilon$.
One further sees that larger values of $\epsilon$ tend to suppress mostly the outer radius $r_+$
whereas the inner radius $r_-$ experiences only modest changes.

\begin{figure}[hbt]
   \centering
\includegraphics[width=10cm]{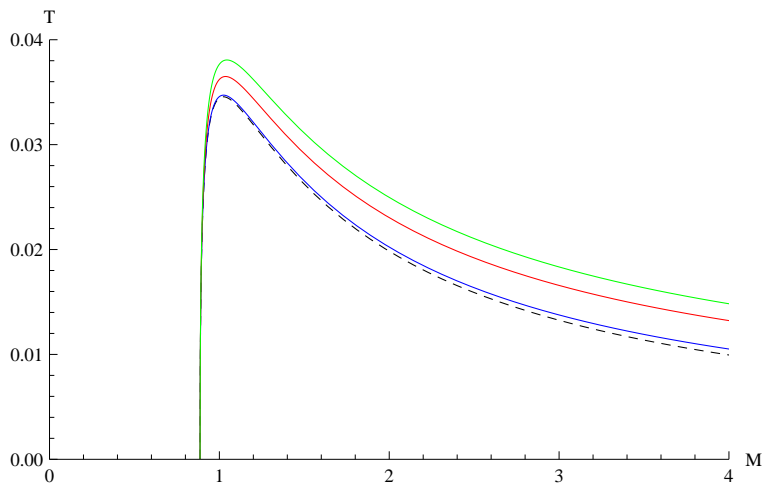}
  \caption{Temperature curve as a function of black hole mass parameter 
  $M_0$ for $G_0=1$, $e^2_0=1$, $Q_0=0.25$, and $\epsilon = \{0,0.1, 0.5, 0.8\}$,
 plotted in black-dashed, blue, red, and green respectively. Due to the absence of 
 first order corrections, the values of the correction parameter $\epsilon$ had to be chosen significantly
 larger than in the previous figures.}
\label{fig:TMRN}
\end{figure}

This outer horizon is responsible for the thermodynamic behavior of the black hole.
Imposing regularity around this point one obtains the standard temperature by
\begin{equation}
T=\frac{\partial _{r} f(r)}{4 \pi} \biggr \rvert _{r_{+}}\quad.
\end{equation}
Evaluating this for the solution (\ref{solRN}) one obtains
\begin{equation}
T=\frac{\epsilon ^2  \left(2 \epsilon  G_{0} M_{0} \left(\sqrt{G_{0} \left(G_{0} {M_{0}}^{2} -\frac{4 \pi 
   {Q_{0}}^{2} }{{e_{0}}^{2}}\right)}+{G_{0}} {M_{0}}\right)+\sqrt{{G_{0}} \left({G_{0}} {M_{0}}^{2} -\frac{4 \pi 
   {Q_{0}}^{2} }{{e_{0}}^{2} }\right)}-8 \pi  \epsilon  {G_{0}}\frac{ {Q_{0}}^{2}}{{e_{0}}^{2}} \right)}
   {2 \pi    \left(2 \epsilon  \left(\sqrt{{G_{0}}
   \left({G_{0}} {M_{0}}^{2} -\frac{4 \pi  {Q_{0}}^{2} }{{e_{0}}^{2} }\right)}+{G_{0}} {M_{0}}\right)+1\right)^{3/2} \left(\sqrt{2 \epsilon 
   \left(\sqrt{{G_{0}} \left({G_{0}} {M_{0}}^{2} -\frac{4 \pi  {Q_{0}}^{2} }{{e_{0}}^{2} }\right)}+{G_{0}} M_{0}\right)+1}-1\right)^2}\quad.
\end{equation}
This temperature in function of the mass parameter $M_0$, is shown in figure \ref{fig:TMRN}.
In order to get somewhat more analytical insight on this cumbersome expression one can expand it for small values of $\epsilon$
\begin{eqnarray}\label{TRNapprox}
T = \frac{\sqrt{{G_{0}} \left({G_{0}} {M_{0}}^{2} -\frac{4 \pi  {Q_{0}}^{2} }{{e_{0}}^{2} }\right)}}{2 \pi  \left(\sqrt{\frac{G_{0} \left({e_{0}}^{2} 
   {G_{0}} {M_{0}}^{2} -4 \pi  {Q_{0}}^{2} \right)}{{e_{0}}^{2} }}+{G_{0}} {M_{0}}\right)^2} + 
    \frac{\epsilon ^2 G_0\left( {G_{0}}{M_{0}}^{2} 
    + M_{0} \sqrt{\frac{{G_{0}} \left({e_{0}}^{2}{G_{0}}
  {M_{0}}^{2} -4 \pi  {Q_{0}}^{2} \right)}{{e_{0}}^{2}}}-4 \pi  \frac{{Q_{0}}^{2}}{e_0^2} \right)}
  {8 \pi  \left(\sqrt{\frac{{G_{0}}
   \left({e_{0}}^{2} {G_{0}} {M_{0}}^{2} -4 \pi  {Q_{0}}^{2} \right)}{{e_{0}}^{2} }}+{G_{0}}{M_{0}}\right)} +O(\epsilon ^{3})\quad.
\end{eqnarray}
The first term of this expansion corresponds to the expected classical limit,
which corresponds to the black dotted line in figure \ref{fig:TMRN}.
The second term of this expansion is turns out not be linear in the expansion parameter
but rather to order $\epsilon^2$, indicating that corrections to the classical temperature tend to be
suppressed for small $\epsilon$.
One further sees from the second term in (\ref{TRNapprox}), that first corrections to the temperature 
are expected to be positive.
For a given $G_0,\, M_0, Q_0$, and $e_0^2$ this means that the classical temperature is the minimal temperature
found under a variation of $\epsilon$.
Those observations can be readily confirmed by the behavior of the curves in figure \ref{fig:TMRN}.

\section{Summary and Conclusion}
\label{sec_sum}

This paper presents and studies two black hole solutions of the Einstein-Hilbert and Einstein-Maxwell equations,
generalized to the case of scale dependent couplings.
The usual ambiguity due to model dependence of the functional form of those couplings 
($\{G_k,\, \Lambda_k \}$ and $\{G_k, 1/e_k^2\}$ respectively)
is circumvented by promoting the couplings to fields in the equations of motion
($\{G(r),\, \Lambda(r) \}$ and $\{G(r), 1/e(r)^2\}$ respectively). 
The resulting mismatch between unknown functions and independent equations of motion
is absorbed by taking the common ansatz for the spherically symmetric metric field 
$g^{00}\sim 1/g^{11}$, which can be motivated by the
known form of the classical solutions (solutions with scale independent couplings).

The findings for the generalized solution of the Einstein-Hilbert case (\ref{GrEH}-\ref{LrEH}) are:
\begin{itemize}
\item  The solution presents two additional arbitrary constants with respect to the 
classical  (anti) de Sitter-Schwarzschild black hole ($\epsilon$ and $c_4$).
Those constants can be chosen such, that they produce a well behaved
classical limit in the sense that one of the additional constants of integration ($\epsilon$) 
parametrizes deviations from the classical solution.
This implies that in the limit of $\epsilon \rightarrow 0$
the classical  (anti) de Sitter-Schwarzschild black hole is recovered
\item The asymptotic behavior for small radial coordinate shows that the 
classical singularity persists for the generalized solution
\item The asymptotic behavior for large radial coordinate shows that
two of the additional constants of the generalized solution ($c_4$ and $\epsilon$) can produce
a shifted value of the classical value of the cosmological constant $\Lambda_0 \rightarrow \tilde \Lambda$.
The shift disappears for $c_4=1/\epsilon$
\item The numerical and the perturbative study of the horizons of the solution reveals that
the scale dependence parameter $\epsilon$ tends to reduce the value of the
inner and of the cosmological horizon. 
For the  de Sitter case, the mass value of the extremal black hole tends to increase with $\epsilon$
\item The numerical and the perturbative study of the radiation behavior of the generalized solution
reveals that the scale dependence produces a slight increase in the temperature, which is 
only of relative importance for the largest mass values close to the critical value
\end{itemize}

The findings for the generalized solution of the Einstein-Maxwell case (\ref{solRN}) are:
\begin{itemize}
\item The generalized solution has one additional constant of integration with respect to the
classical Reissner-Nordstr\"om black hole. The additional constant of integration (again labeled $\epsilon$) 
can be chosen such that it allows to recover the classical result for the limit $\epsilon \rightarrow 0$
\item The asymptotic behavior for small radii reveals that the singularity at
the radial origin persists (actually it becomes even more visible since $R\neq 0$)
\item The asymptotic behavior for large radii shows that the solution does approach a cone-like asymptotic
space-time, similar (but not identical) to the  asymptotics of known black hole monopoles  \cite{Barriola:1989nn,Shi:2009nn}.
By integrating Gauss' law for this asymptotic (cone-like) space-time one observes that
the effects of the monopole cancel and the 
resulting charge resembles the classical charge parameter $Q=Q_0/e_0^2$, independently
of the value of $\epsilon$
\item The study of the horizon structure of the generalized solution can be 
performed exactly without the need of an expansion in $\epsilon$.
One finds that
the two physical horizons are shifted towards smaller values with respect to the two classical horizons.
This shift turns out to be $\epsilon$ dependent and more important for the outer horizon $r_+$.
Surprisingly this $\epsilon$ dependence cancels out when one evaluates the critical black hole mass (\ref{CosCens1}),
implying that the classical ``cosmic censorship'' relation remains unchanged, independent of the scale
dependence parameter $\epsilon$
\item The thermodynamic behavior of the generalized solution is calculated, showing that 
the scale dependence parameter produces a slight relative increase of the temperature with respect
to the classical solution. This observation is confirmed by a numerical and a perturbative analysis.
\end{itemize}

In summary, the analysis of the presented solutions reveals that scale dependence
of the couplings ($\{G(r),\, \Lambda(r) \}$ and $\{G(r), 1/e(r)^2\}$ respectively),
can not be expected to resolve the problem of singularities at the origin,
but it can produce important effects on the asymptotic space-time
resulting either in a modified cosmological constant (\ref{tildeL2}) or in
an asymptotic monopole (\ref{monop1}).
Effects on the critical masses 
and thermodynamic behavior are either rather mild or even absent as in the case of (\ref{CosCens1}).

\section*{Acknowledgements}
Thanks to Betti Hartmann for suggesting the RN calculation.
The work of B.K.\ was supported proj.\ Fondecyt 1120360
and anillo Atlas Andino 10201.
P.R.\ was supported by beca Conicyt and proj.\ Fondecyt 1120360.







\end{document}